\documentclass{emulateapj}

\shorttitle{Filaments in galactic winds}
\shortauthors{Ary Rodr\'\i guez-Gonz\'alez et al.}

\begin{document}

\title{Filaments in galactic winds driven by young stellar clusters}

\author{A. Rodr\'\i guez-Gonz\'alez\altaffilmark{1},
A. Esquivel\altaffilmark{1}, P. F. Vel\'azquez\altaffilmark{1}, 
A. C. Raga\altaffilmark{1}, V. Melo}
\altaffiltext{1}{Instituto de Ciencias Nucleares, Universidad
Nacional Aut\'onoma de M\'exico, Ap. 70-543,
04510 D.F., M\'exico}

\email{ary, esquivel, pablo, raga@nucleares.unam.mx, vmelo@iac.es}

\begin{abstract}
The starburst galaxy M82 shows a system of H$\alpha$-emitting filaments
which extend to each side of the galactic disk. We model these filaments
as the result of the interaction between the winds from a distribution
of Super Stellar Clusters (SSCs). We first derive the condition necessary
for producing a radiative interaction between the cluster winds (a condition
which is met by the SSC distribution of M82). We then compute 3D simulations
for SSC wind distributions which satisfy the condition for a radiative
interaction, and also for distributions which do not satisfy this condition.
We find that the highly radiative models, that result from the
interaction of high metallicity cluster winds, produce a structure of H$\alpha$
emitting filaments, which qualitatively agrees with the observations of
the M82, while the non-radiative SSC wind interaction models do
not produce filamentary structures.  Therefore, our criterion
for radiative interactions (which depends on the mass loss rate and the
terminal velocity of the SSC winds, and the mean separation between SSCs)
can be used to predict whether or not an observed galaxy should have
associated H$\alpha$ emitting filaments.
\end{abstract}

\keywords{Hydrodynamics -- shock waves -- star: winds, outflows
 -- galaxies: star cluster -- galaxies: starburst -- 
galaxies: individual: (M82, NGC253)}

\section{Introduction}

Supergalactic winds are a very important mechanism 
in the evolution of the Universe. Lynds \& Sandage (1963) found 
a large outflow in M82 and gave the first 
definition of a starburst. They proposed that material was ejected from
the nuclear regions as the consequence of a SN explosion. Different
analytical  models have been put forward to explain the so-called
supergalactic winds (SGWs), see, e. g., the work of
Chevalier \& Clegg (1985), Tomisaka \& Ikeuchi  (1988), Heckman et al
(1990), Leitherer \& Heckman (1995), Strickland \& Stevens (2000), Cant\'o
et al (2000), Tenorio-Tagle et al. (2003), and Cooper et al. (2008). 
In the nearest starbursts, they have shown that the supergalactic winds (SGW)
form through the collective effects of many individual stellar cluster winds, 
which are in turn  formed by the collective effect of many individual 
stellar winds.

The best-studied starburst galaxy is M82. It has an extended, biconical
filamentary structure in the optical (Shopbell \& Bland-Hawthorn 1998;
Ohyama et al. 2002).
This optical emission is embedded in a pool
of soft  X-ray emission, detected with the Chandra X-Ray Observatory
(Griffiths et al. 2000) as well as with XMM-Newton  (Stevens et
al. 2003). This X-ray emission extends several kpc away from the
nuclear region, into the IGM. 

With the Hubble Space Telescope (HST), it has been possible to isolate
the building blocks of some starbursts:Super Stellar Clusters
(SSCs). This objects are very massive 
and dense stellar clusters, with masses
in excess of $10^4~\rm{M_{\odot}}$ enclosed in radii of $3$--$10$
pc. They are young, and typically contain from $\sim 100$
up to $\sim 10^4$ OB stars. 
In M82, the SSCs $H{\alpha}$ luminosities are in the range of
(0.01-23)$\times 10^{38}~\rm{erg~s^{-1}}$  (Melo et al 2005). 
These authors  catalogued 197 SSCs in M82 with masses in 
the $10^4< \rm{M/M}_{\odot}<10^6$ range. This exceptional density of
massive clusters ($620~\rm{kpc}^2$) produces a suitable scenario 
for a study of the interaction between cluster winds.

The differences in the populations of SSCs of
several starburst  galaxies can help us to understand which properties
are more important for producing filaments.
For instance  NGC253, which does not show a clear filmentary structure
(Rice 1993), has a significantly lower SSC mass loss rate than 
M82.
However, other authors claim 
that the lack filamentary structure in this galactic wind is simply a
result of not having deep enough H$\alpha$ observations of this
objects (Hoopers et al. 1995).   In NGC253 Melo (2005) catalogued a total of 
48 SSCs with an average distance
between them of $31~\rm{pc}$, in contrast with the mean separation between
SSCs of $12~\rm{pc}$ found in M82. In M82, a rich structure
of filaments, extending out from the galactic plane, is present. There
are many other examples such as NGC1569 
(Anders et al. 2004; Westmoquette, Smith \& Gallagher III 2008) and
 M83 (Harris et al. 2001), that have high density of
SSCs with similar masses to the ones of M82. 

Tenorio-Tagle et al. (2003) showed that the interaction 
of stellar winds from a collection of energetic, neighbouring SSCs
could form a filamentary structure similar to the one observed in M82.
These authors propose the formation of such filaments by means of
supernovae explosions that would expel a huge amount of heavy elements into
the Interstellar Medium (ISM), thus enhancing the radiative cooling of
the outflows.

In their model, Tenorio-Tagle et al. (2003) explained the SGW
structure as a  consequence of the interaction of winds from very
close SSCs, in which stationary oblique shocks
are responsible for shaping  the material into dense and cold,
kiloparsec-sized  filaments. More recently, Cooper et al. (2008)
presented a numerical study of a starburst-driven galactic wind. Their
setup consisted of  a series of neutral dense clouds placed in the 
galaxy disk. The dense clouds are swept up by the main
shock wave, leaving behind a filamentary structure.

In the present paper, we discuss models for the formation of
filamentary structures as the result of the interaction between
winds from a system of many clusters with a disk-like spatial
distribution. In these models we consider only the flow resulting
from the wind interactions, and do not include the effect of the
stratified ISM present in the region of the galactic disk.

The paper is organized as follows. In \S 2, we study which combinations
of parameters (mass loss rate ${\dot M}_w$, cluster wind velocity $v_w$,
and separation between nearby star clusters $D$) yield a highly
radiative SGW flow. In \S 3 we describe a series
of 3D simulations of the interaction of
cluster winds in this highly radiative regime.
We present predictions of the emission in the optical and X-ray
wavelength ranges, and
compare them with observations of galactic winds (i.e. M82, NGC253,
etc.) in \S 4. Finally, in \S 5 we present our conclusions.

\section{The formation of filamentary structures}

\subsection{General considerations}

The collective effect of the individual stellar winds inside an SSC
result in a cluster wind. Outside the outer radius of the cluster (i. e., for
radii $r>r_c$, where $r_c$ is the cluster radius), the cluster
wind has an approximately constant velocity, and a density that
falls $\propto r^{-2}$ (where $r$ is the
distance to the cluster centre, see, e.~g.,  Rodr\'iguez-Gonz\'alez et
al. 2007).

Cold filaments in SGWs can be formed by the the interaction of such
SSC winds, provided that the cooling is efficient.
Since the gas density of the cluster wind falls  with
distance, this is likely to happen only when the clusters are very
close each other. The terminal velocity of the
SSC winds and their mass loss rate are also important for determining
whether or not we have efficient cooling.

\subsection{Radiative losses in a cluster wind}
\begin{deluxetable}{cccc}
\tabletypesize {\scriptsize}
\tablecaption{Cooling distance as a function of shock velocity}
\tablewidth{0pt}
\tablehead{
\colhead{v$_{s}$ [km s$^{-1}$]} &  & \colhead{log$_{10}$(d$_{c,1}$
  [pc])}&  \\
\cline{2-4}  
 & \colhead{Z$_{\odot}$ }& \colhead{5Z$_{\odot}$}&  \colhead{10Z$_\odot$}
}
\startdata
100    & -1.468& -2.010& -2.283\\
200    & -1.194& -1.960& -2.259\\
300    & -0.657& -1.439& -1.796\\
400    &  0.595& -0.091& -0.399\\
500    &  0.957&  0.269& -0.037\\
600    &  1.176&  0.516&  0.209\\
700    &  1.449&  0.763&  0.455\\
800    &  1.601&  0.913&  0.605\\
900    &  1.727&  1.038&  0.714\\
1000   &  1.893&  1.208&  0.900\\
1100   &  2.108&  1.438&  1.123\\
\enddata
\end{deluxetable}
Let us consider a galaxy with a central stellar cluster density $n$ (number
of clusters per unit volume). For the sake of simplicity, we consider
stellar clusters with identical, isotropic winds,  mass loss rate
${\dot M}_w$, and a terminal wind velocity $v_w$.
The typical separation between stellar clusters is $D\approx n^{-1/3}$.

Therefore, two-wind shock interactions between nearby clusters pair will
occur at a typical distance $\sim D/2$ from each of the stellar clusters,
so that the pre-shock densities would have values of
\begin{equation}
n_{pre}={{\dot M}_w\over 1.3 m_H\pi D^2 v_w}\,,
\label{npre}
\end{equation}
where $m_H$ is the Hydrogen mass, and we have assumed a 90\%\ H and
10\%\ He particle abundance.

The shock interactions between nearby clusters will be radiative if
the cooling distance $d_{cool}$ satisfies the condition
\begin{equation}
\kappa \equiv {d_{cool}\over D}<1\,.
\label{k}
\end{equation}

In order to estimate the cooling distance, we proceed as
follows. Consider the structure that will be formed by the leading
shock of a star cluster wind with velocity $\approx v_w$ (i. e. the
cluster wind velocity) with  a preshock density given by equation
(\ref{npre}). For a cooling function in the low density regime,
the cooling distance scales as the inverse
of the pre-shock density, yielding:
\begin{equation}
d_{cool}(n_{pre},v_s)=\left[{{d_{c,1}(v_s)}\over n_{pre}}\right],
\label{dc}
\end{equation}
where $v_s$ ($=v_w$) is the shock velocity, and $d_{c,1}(v_s)$ is the
cooling distance (to $10^4$~K) behind a shock with $n_{pre}=1$~cm$^{-3}$.
We have calculated the function $d_{c,1}(v_s)$ by integrating the
equations for a 1D, stationary shock (see, e.~g., Shull \& McKee 1979)
using the coronal equilibrium cooling functions for 1, 5 and 10 
times solar metallicities of Raymond, Cox \& Smith (1976). The
resulting values of the cooling distance $d_{c,1}(v_s)$ (to $10^4$~K,
for a $n_{pre}=1$~cm$^{-3}$ preshock density) are given in
Table 1.

Setting $\kappa=1$ in equation (\ref{k}) and also using
equation (\ref{dc}), one obtains
\begin{equation}
\frac{\dot{M}}{D}=1.3 m_H \pi v_w d_{c,1}(v_w)\,,
\label{frac}
\end{equation}
which gives the relation between ${\dot M}$, $D$ and $v_w$ for
SSCs that are just entering the highly radiative regime.
\begin{figure}
\centering
\includegraphics[width=9.7cm]{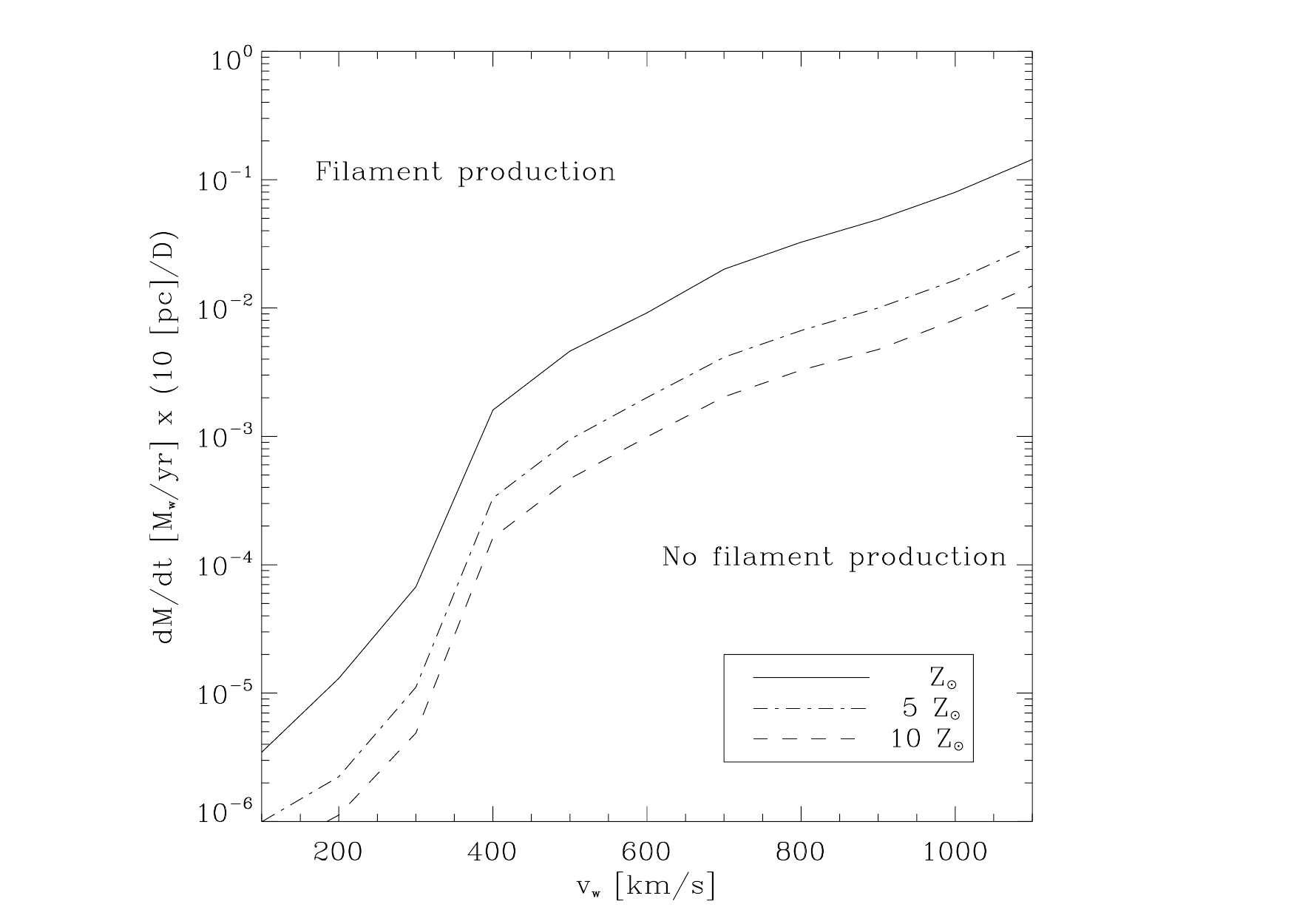}
\caption{This diagram shows
${\dot M}_w \times$ (10 pc/D) as a function of $v_w$
(where $D$ is the separation between clusters,
${\dot M}_w$ is the mass loss rate, and $v_w$ the cluster wind
velocity) for a ``cooling parameter'' $\kappa=
d_{cool}/D=1$. Three curves are shown, corresponding to models
with 1, 5, and 10 times solar metallicities.
The region above the curves represents the parameter space
in which the two-wind interactions of cluster winds are radiative.} 
\label{cool}
\end{figure}
In Figure 1, we plot the $\frac{\dot{M}}{D}$ ratio
as a function of $v_w$ and for SSCs with $\kappa=1$ (see equation~\ref{frac})
for the three metallicities which we have considered (see above and Table~1).
The curves that would be obtained for $\kappa<1$ always lie above
the $\kappa=1$ curve corresponding to the same metallicity.
Figure 1 shows that in a substantial part of the parameter space
which would be expected for galaxies with low or intermediate mass
young stellar clusters, cooling parameters $\kappa<1$ are produced.
A galactic wind in this ``highly radiative regime'' will have dense, cool
structures resulting from the radiative shocks in the interactions
between nearby stellar clusters. In the following section, we present
numerical simulations of flows in this regime.

\section{Numerical simulation}
We have computed 3D numerical simulations with the full, radiative
gasdynamic equations. The simulations solve a multiple stellar
wind interaction problem with the 3D, adaptive grid 
``yguaz\'u-a'' code, which is described in detail by Raga et al. (2000,
2002). The first four simulations were performed on a
five-level binary adaptive
grid with a maximum resolution of 0.488 pc (corresponding to
512$\times$512$\times$1024 grid points at the maximum grid resolution),
in a computational domain of 250$\times$250$\times$500 pc.
We have also computed a fifth simulation also on a five-level
adaptive grid, but with a maximum resolution of 0.976 pc, corresponding
to $1024 \times 1024 \times 2048$ grid points (at the maximum
grid resolution) in a domain of $1\times 1 \times 2$~kpc (see Table 2).

In order to study the formation of filaments in the radiative
regime we ran 6 models. We have
assumed that the computational domain was filled initially with a
homogeneous, stationary ambient medium with temperature 
${T_{env}= 1000~\rm{K}}$ and density ${n_{env}=0.1~\rm{cm}^{-3}}$.
This environment clearly does not include the ISM of the disk
of a starburst galaxy, for which densities of $\sim 1000$~cm$^{-3}$
might be more appropriate. We choose this low density so that the
ISM in the region in between the cluster is rapidly flushed
away by the cluster winds, and a stationary wind interaction
structure is reached as rapidly as possible. If we introduced
a higher ISM density (within the galactic plane), there would
be a transient regime (lasting longer for higher
ISM densities) before the cluster winds break out of the galactic
plane and travel into the intergalactic medium. The density which
we have chosen is closer to a possible density for the inter galactic
medium into which the cluster winds will be expanding once they
leave the galaxy.

We model the SSCs as spherical wind sources, centered at the positions
as described below.
For models M1-M4 (see Table 2), the wind sources have a radius of
$R_c=2.92$ pc (corresponding to 6 pixels at the maximum resolution of
the adaptive grid, which is always present at the wind sources),
for models M2b and M3b their radii is $R_c=5.84$ pc 
(also corresponding to 6 pixels at the highest resolution grid).

Within these spheres, we impose (at every timestep) a wind with a
temperature of $T_c=10^7$~K, an outwardly directed velocity
$v_c=1000~\rm{km~s}^{-1}$, and a density that follows an $r^{-2}$ law
(where $r$ is the radial coordinate measured outwards from the position
of each wind source), scaled so as to obtain the correct the mass loss rate.

The positions of the wind sources are obtained by sampling a uniform
random distribution. However, in practice, we have to modify the
obtained cluster distributions to avoid the overlap of sources.
The stellar cluster sources are placed in a cylindrical
structure, meant to model the central region of a galactic disk.
 \begin{table}
\begin{center}
\caption{Model parameters}
\begin{tabular}{ccccc}
\tableline
\tableline
Model & Number of & $Z$         & $D$\tablenotemark{\dag}& $\dot{M}_{SC}$\\
      & SSCs      &$[Z_{\odot}]$& [pc] & [$\rm{M}_\odot\,\rm{yr}^{-1}$]\\
\tableline
M1   &  100      &   1          &  9.27   &$2\times10^{-2}$\\
M1a  &  100      &   1          &  9.27   &$2\times10^{-1}$\\ 
M2\tablenotemark{\ddag}&  100   &   5   &  9.27   &$2\times10^{-2}$\\
M3\tablenotemark{\ddag}&  100   &   10  &  9.27   &$2\times10^{-2}$\\ 
M4   &  15       &   10         &  21.88  &$4\times10^{-3}$\\
M4a  &  15       &   10         &  21.88  &$4\times10^{-2}$\\
\tableline
\end{tabular}
\tablenotetext{\dag}{The value quoted here is the mean distance between
 neighbouring clusters (nearest neighbors) as measured from the 
positions that result from sampling a uniform random distribution.} 
\tablenotetext{\ddag}{We have also run models M2b and M3b with the same
parameters as M2 and M3, but with 1/2 of the spatial resolution
and 4 times the spatial extent as M2 and M3.}
\end{center}
\end{table}
\subsection{The starburst regions}

In order to reproduce the  properties of M82 and of NGC253, we use the
average properties reported  by Melo et al. (2005) 
and Melo (2005). For M82, the mean stellar cluster mass ($\bar{M}_{SC}$) is
of $1.7\times10^5~\rm{M}_\odot$, and the mean separation between
neighbouring clusters ($\bar{\Delta}$) is of 12 pc.
For NGC253, $\bar{M}_{SC}$=6.2$\times$10$^4$ M$_\odot$ and
$\bar{\Delta}$=31 pc. We should note that the clusters of Melo et al.
(2005) were detected at optical wavelengths. As there is a high
level of obscuration in the nuclear regions of M82 as well as NGC253, 
the real number of clusters will be larger, and the mean separation between
clusters will be smaller than the values quoted above.

We estimate the mass loss rate $\dot{M}_{SC}$ using the starburst 99 models
(Heckman \& Leitherer 1995, Leitherer et al. 1999 
\footnote{http://www.stsci.edu/science/starburst99/}) with the appropriate
cluster masses and a Salpeter initial mass function (IMF) including
stars between 1 and 100 M$_\odot$.
The resulting mass loss rate (through winds and supernovae) is
reasonably constant after $\sim 4~\mathrm{Myr}$ of the onset of the
starburst, with some enhancement when input by SNe overcomes the input
from winds.  This enhancement occurs at around $t\simeq 10~\mathrm{Myr}$, 
and lasts for about  $10~\mathrm{Myr}$, then the mass loss rate
returns to a value similar to the pre-supernovae phase for another
$\sim 20~\mathrm{Myr}$. Finally, the mass loss rate drops
dramatically once the last star of  $\sim 8~\rm{M}_\odot$ explodes as
a supernova leaving only intermediate and low-mass star winds (at this
point the star cluster would no longer be classified as SSC).


\begin{figure*}[!ht]
\centering
\includegraphics[width=12cm]{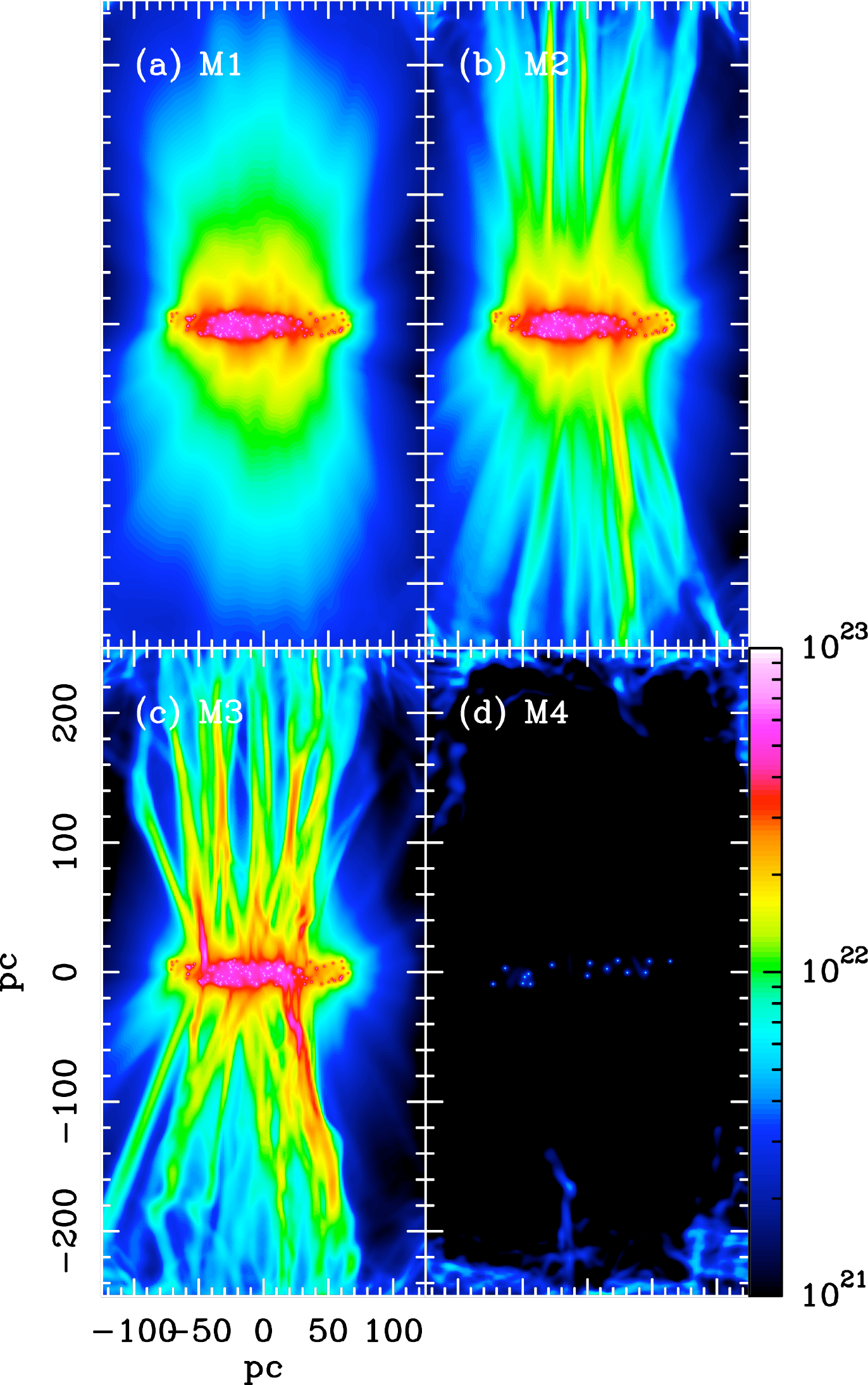}
\caption{Column density maps of models M1-M4. The bar at the bottom
  right gives the logarithmic gray-scale (color-scale in the online
  version) in $\rm{cm}^{-2}$. The images were obtained
  by integrating density along lines of sight
  parallel to the $x$-axis. The distributions correspond to
  time-integrations of $(2.5,2.5,2.5,5)\times 10^5$~yr (for models
  M1-M4, respectively).}
\end{figure*}

For models M1-M4 we used a constant mass loss rate with a value
consistent with the average mechanical luminosity of the SSCs.
However, given the assumptions we have made (of having a system of identical 
clusters with average properties) a larger mass loss rate could be possible,
especially if the SSCs are in the SN phase. Such a large mass loss rate
could have a significant effect in the production of filaments.
For this reason we have run two additional models, M1a and M4a, with
extreme values of $\dot{M}_{SC}$, which correspond to upper limits
(within the uncertainties), at the peak of the mass loss rate durinng
the SN phase. Aside of the increased mass loss rate, these two models
are identical to M1 and M4, respectively (see Table 2).
\begin{figure*}
\centering
\includegraphics[width=12cm]{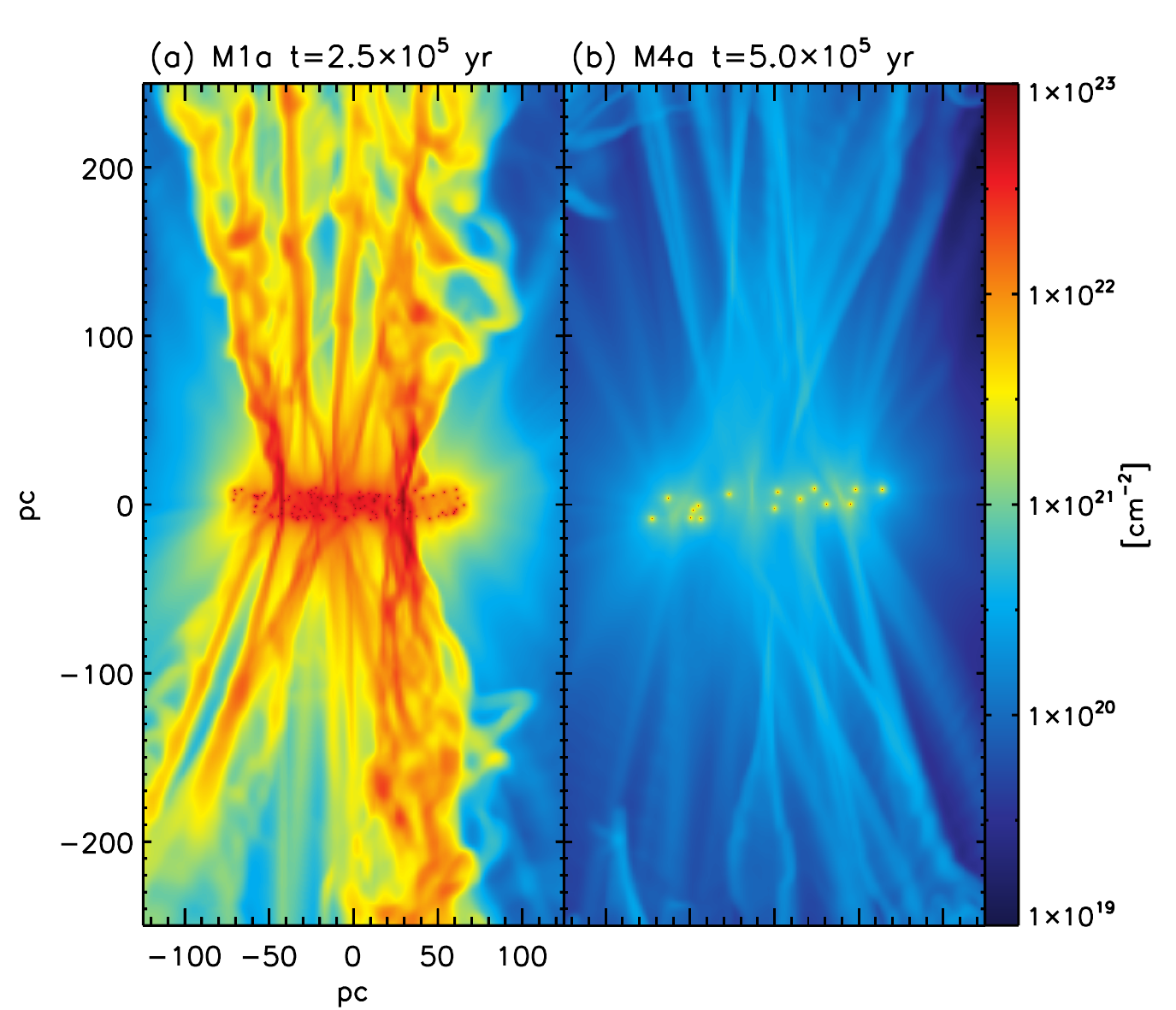}
\caption{Column density maps for the models with mass loss rate 
in the SN phase (M1a and M4a). The bar at the   right gives the 
logarithmic gray-scale  (color-scale in the online  version) in 
$\rm{cm}^{-2}$. The images were obtained  by integrating the density along 
lines of sight parallel to the $x$-axis.   The corresponding
integration times are of $(2.5,5)\times 10^5$~yr (for models  M1a and
M4a, respectively).} 
\end{figure*}
\subsection{The models}

We computed four numerical simulations of SGWs driven by stellar
clusters, with different metallicities, mass input rates, and number of
clusters. All of the models were computed with identical stellar
clusters with a cluster wind
velocity $v_w=1000~\rm{km~s}^{-1}$, placed in a cylinder with
a radius of $100~\rm{pc}$ and a height of $20~\rm{pc}$.
This cylinder represents the central part of the starburst
in a galaxy as M82, with dimensions approaching those of an individual 
starburst clump (O'Connell et al. 1995, Westmoquette et al. 2007). In reality
the starburst of M82  has spatial extent of $\sim 500$~pc (see O'Connel 
\& Mangano 1978). Numerical simulations with the full spatial extent for the
cluster distribution would be much more demanding computationally.
From SB99, using the mean cluster mass for M82 (Melo 2005), we 
estimate a mechanical luminosity  of $6\times
10^{39}~\rm{erg~s^{-1}}$ per SSC. Using $\dot{E}_{SC}=\dot{M}_{SC}\,v_w^2/2$
relation, this luminosity translates into a cluster mass loss  rate of 
$\sim 2\times 10^{-2}~\rm{M_{\odot}~yr^{-1}}$. This cluster mass loss rate
has been used for computing models M1, M2, and M3.
These three models have different metallicities~:
$Z=1,~5$ and $10 Z_\odot$ (for M1, M2 and M3 respectively).

These metallicities have been chosen as representative values
of the metallicities of winds from starbursts during the first
20 Myr time-evolution. Using the metal yields of Meynet
\& Maeder (2002), Tenorio-Tagle et al. (2005) have shown
that the metallicity of the combined stellar winds remains
at a value of $\sim 0.5\, Z_\odot$ up to a $t=3$~Myr
evolutionary time, rapidly grows to $\sim 15\, Z_\odot$
at $t\approx 7$~Myr, and then gradually decreases to
$\sim 3\, Z_\odot$ at $t\approx 20$~Myr.

Similarly, for NGC253  we obtain a mechanical luminosity of
$1.9\times 10^{39}~\rm{erg~s^{-1}}$, which corresponds to a cluster
mass loss rate $\sim 4\times 10^{-3}~\rm{M_{\odot}~yr^{-1}}$. This
cluster mass loss rate was used for computing model M4.

For models M1-M3, we have placed 100 clusters within the cylindrical
volume described in the beginning of this subsection, which results
in a mean separation between clusters $D=9.3$~pc, which is similar
to the mean separation $\bar{\Delta}=12$~pc between the clusters
of M82 (see \S 3.1). Model M4 has a $Z=10 Z_\odot$ metallicity,
and has 15 clusters within
the cylindrical volume, resulting in a mean separation between
clusters of $D=21.9$~pc, which is similar
to the mean separation $\bar{\Delta}=31$~pc between the clusters
of NGC253 (see \S 3.1). The parameters of models M1-M4 are summarized
in Table 2.

With these parameters, models M2 and M3 lie above the
$\kappa=1$ curves (for the appropriate metallicities), so
that they are clearly in the highly radiative regime (see Figure 1).
Model M1 and M4 lie below the $\kappa=1$ curves, so that they
are not in this regime. Therefore, we would expect to form dense
filaments only in models M2 and M3.
We evolve models M1, M2, and M3 up to an integration time of $t=2.5\times
10^{5}~\rm{yr}$, and M4 up to $t=5\times 10^{5}~\rm{yr}$.
In Figure 2, we show the column densities obtained by integrating
the number density along the $x$-axis of the domain, for the flow
stratifications obtained at the end of the time-integrations. From
this Figure, it is clear that models M2 and M3 produce a structure of
$\sim 10\to 20$ filaments both above and below the plane of the
galactic disk (within which are distributed the SSCs). Models M1 and
M4 do not produce filamentary structures.

The bipolar distribution of the filaments is a direct result of the
flattened cylindrical distribution of the SSCs (see Figure 2). We
have also done simulations in which the cylindrical distribution
has a height equal to the diameter (not included in the present
paper), and they show filamentary structures with an almost isotropic
distribution. The same result is of course also found for a spherically
symmetric cluster distribution.

In Figure 3, we present the column density obtained in the models with
extreme mass loss rates (M1a and  M4a, see Table~2) at t=2.5 and 5
$\times 10^5$~yr, respectively.
Due to the increase of $\dot{M}$, the cooling distances become an
order of magnitude shorter (see the density scaling in equation 3),
thus the adimensional parameter $\kappa$ for M1a and M4a is comparable
to that of M3 ($\kappa < 1$).
Therefore, as expected, Figure 3 shows filamentary scructure, even
when it is absent for the same configuration at a lower mass loss. 
\begin{figure*}
\centering
\includegraphics[width=12cm]{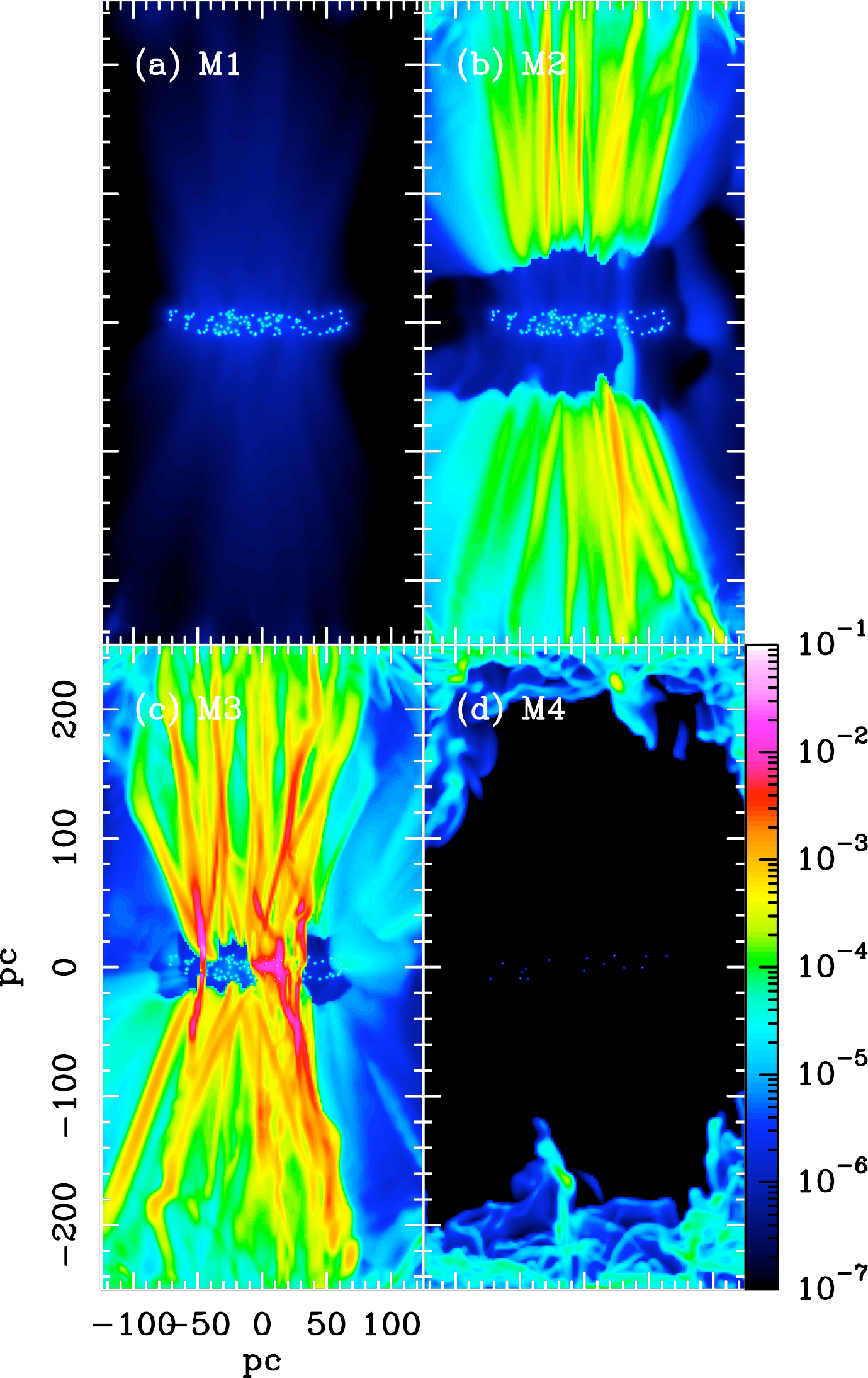}
\caption{H$\alpha$ maps obtained for models M1-M4. The bar at the bottom
  right, gives the logarithmic gray-scale (color-scale in the online
  version) in ${\rm{erg~s}^{-1}~\rm{cm}^{-2}~\rm{sterad^{-1}}}$.  The
  images were obtained by integrating the emission coefficient along
  lines of sight parallel to the $x$-axis. The distributions correspond to
  time-integrations of $(2.5,2.5,2.5,5)\times 10^5$~yr (for models
  M1-M4, respectively).}
\end{figure*}
\section{The galactic wind emission}

\subsection{The H$\alpha$ emission}

In Figure 4, we show the H$\alpha$ maps computed by integrating the
H$\alpha$ emission coefficient along the $x$-axis. The emission coefficient
is computed with the interpolation formula given by Aller (1987) for
the temperature dependence of the recombination cascade.

In model M3, we
see a structure of H$\alpha$ filaments which extend out from the
galactic plane. In model M2 (which has lower radiative losses due to
its lower metallicity), the H$\alpha$ filaments appear only at a distance
of $\sim \pm 50$~pc from the galactic plane. Models M1 and M4 do not
produce H$\alpha$-emitting filaments.
\begin{figure*}
\centering
\includegraphics[width=12cm]{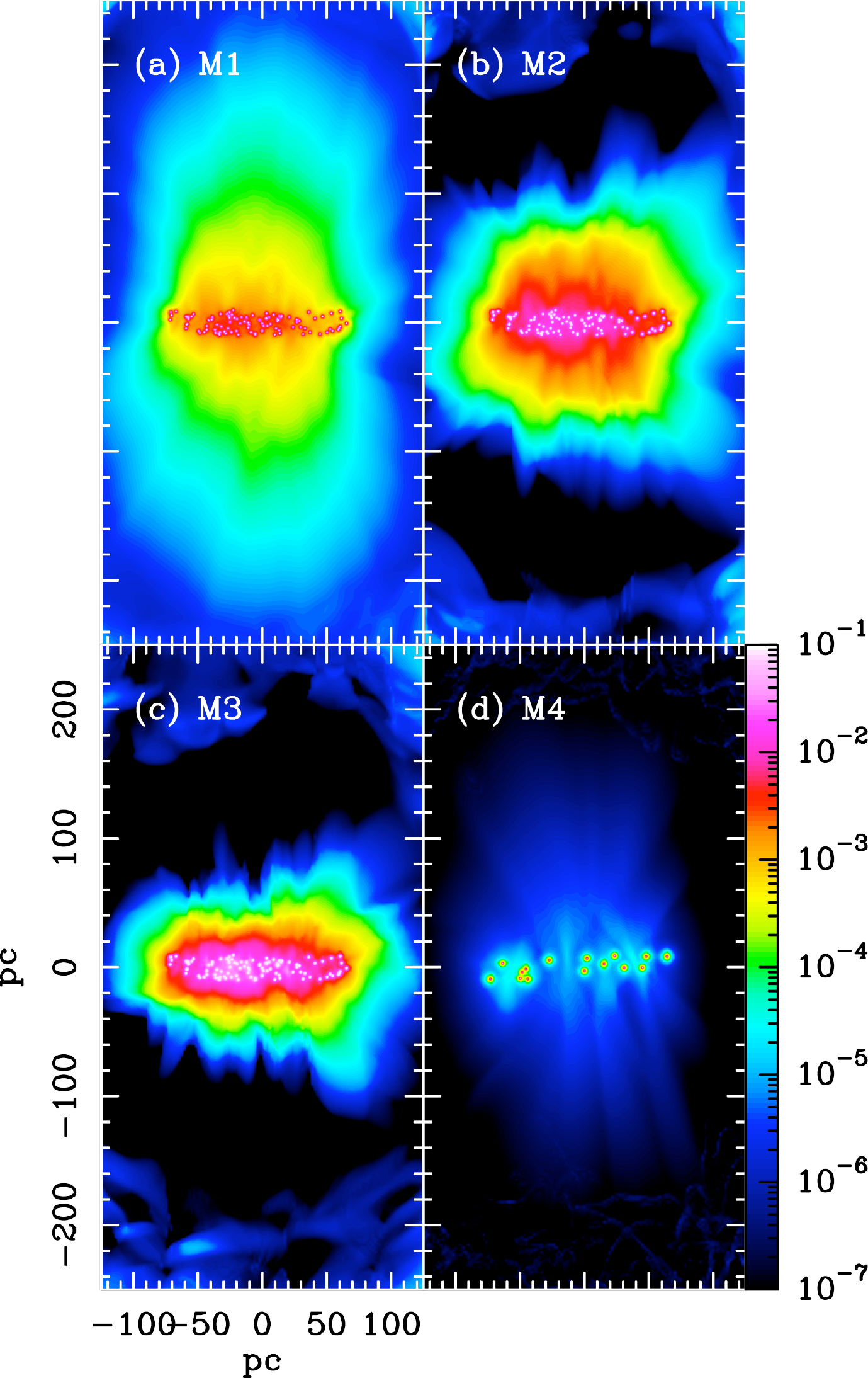}
\caption{X-ray maps for models M1-M4.The bar at the bottom
  right, gives the logarithmic gray-scale (color-scale in the online
  version) in ${\rm{erg~s}^{-1}~\rm{cm}^{-2}~\rm{sterad^{-1}}}$. The
  images were obtained by integrating the 
  X-ray emission coefficient along lines of sight (parallel to $x$-axis),
  in the energy band of $0.3$ to $2$ keV. The distributions correspond to
  time-integrations of $(2.5,2.5,2.5,5)\times 10^5$~yr (for models
  M1-M4, respectively).}
\end{figure*}
The H$\alpha$ filaments in M82 (Ohyama et al. 2002) appear to
extend down to the galactic plane, though this effect could partially
be due to the fact that the plane of M82 is at an angle of $\sim 10^\circ$
with respect to the line of sight. We would therefore conclude that
the $Z=10 Z_\odot$ metallicity model M3 appears to be more appropriate
for reproducing the H$\alpha$ filaments of M82.
\begin{figure*}
\centering
\includegraphics[width=12cm]{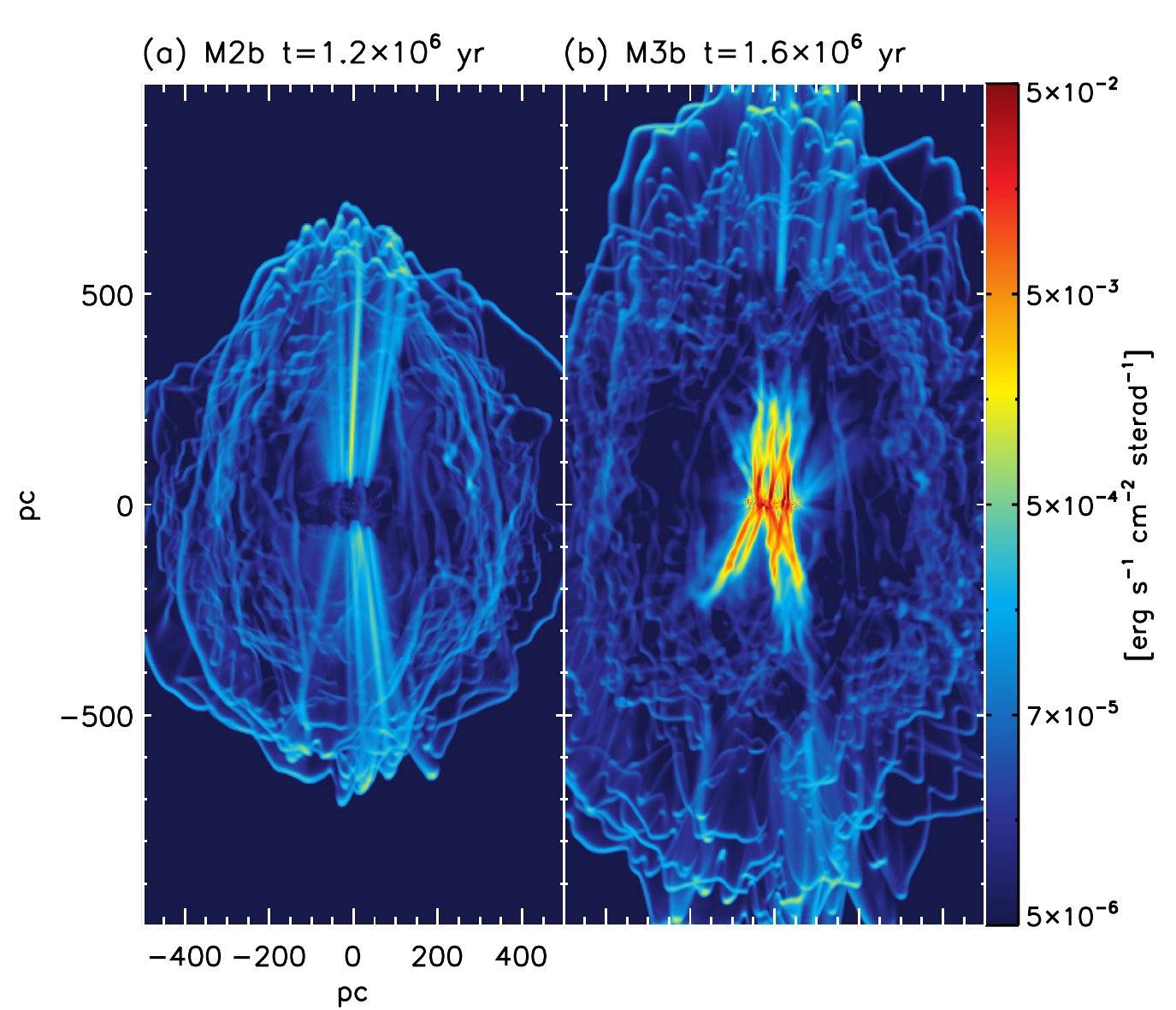}
\caption{H$\alpha$ maps obtained for models M2b (left) and M3b (right).
  The bar on the
  right, gives the logarithmic gray-scale (color-scale in the online
  version) in ${\rm{erg~s}^{-1}~\rm{cm}^{-2}~\rm{sterad^{-1}}}$.  The
  images were obtained by integrating the emission coefficient along
  lines of sight parallel to the $x$-axis. The distributions correspond to
  time-integrations of $(1.2,1.6)\times 10^6$~yr (for models
  M2b and M3b, respectively).}
\end{figure*}
\subsection{The X-Ray emission}

We have taken the density and temperature stratifications from the
SGW flow configurations, and used them to calculate the
X-ray emission. We have done this by computing the emission coefficient
in the $0.3\to 2$~ keV photon energy range using the
CHIANTI~\footnote{The CHIANTI database and associated IDL procedures, 
now  distributed as version 5.1, are freely available at:  
http://wwwsolar.nrl.navy.mil/chianti.html, 
http://www.arcetri.astro.it/science/chianti/chianti.html, 
and http://www.damtp.cam.ac.uk/user/astro/chianti/chianti.html} atomic data 
base and software (see Dere et al. 1997, Landi et al. 2006). For
this calculation, it is assumed  that the ionization state of the gas
corresponds to coronal ionization equilibrium in the low density
regime  (i.~e. the emission coefficient is proportional to the
square of the density).

The X-ray maps corresponding to the end of the time-integrations
of models M1-M4 are shown in Figure~5. These maps show that the
models M1 and M4 (with low radiative losses) produce an extended,
diffuse X-ray emission. On the other hand, the highly radiative models
M2 and M3 produce X-ray emission concentrated to the galactic plane.

M82 shows X-ray emission which extends to large distances (few kpc) from the
galactic plane (Stevens et al. 2003). Because the preferred
model for this galaxy is M3 (which produces H$\alpha$ filaments
which start from the galactic plane, see Figure 4 and \S 4.1), we
see that in order to reconcile the X-ray map predicted from this model
(see Figure 4) and the observed X-ray distribution of M82 we need
to assume that the observed X-ray emission comes from another component.
This component, e. g., could be the shock of the expanding SGW against the
intergalactic medium. The emission from this shock is seen in the
top and bottom parts of the frames corresponding to models M2 and M3
in Figure 5, but the shock structure is already escaping from the
computational domain.

\subsection{The H$\alpha$ emission at larger distances from the galaxy}

Models M1-M4 extend only to 250~pc on either side of the galactic
plane. As the filaments in M82 and NGC1569 extend to $\sim 0.5\to 1$~kpc
away from the plane of the galaxy, we have run models M2b and M3b
(with the same parameters as M2 and M3, respectively, but with lower
resolution and larger spatial extent, see \S 3 and Table 2)
in order to explore whether or not the H$\alpha$ filaments
predicted from our models do extend out to $\sim 1$~kpc.

In Figure 6, we show the H$\alpha$ maps predicted from models
M2b (at $t=1.2\times 10^6$~yr) and M3b (at $t=1.6\times 10^6$~yr).
The two times are chosen so as to show an approximate time
sequence of the flow, as the two models produce similar results.
It is clear that in the two models, the H$\alpha$ filaments
extend to distances of $\sim \pm 500$~pc from the galactic
plane. In the H$\alpha$ map of M3b, there is a transition
to a more complex morphology at distances $>\pm 300$~pc
from the galactic plane, which is not seen in model M2b
(in which basically the same filaments are
seen at larger distances).

\section{Conclusions}

We study the formation of filaments in the galactic winds driven by young
SSCs in regions of high star formation. We study models in which the
filaments are produced solely by the interaction between winds from
SSCs, and do not consider the possible effects of the stratified
or clumped ISM present in the galactic disk. In our models, the 
winds propagate into a low density, homogeneous environment, which 
could correspond to the intergalactic medium.

We have shown how a  $(\dot{M}_w/D)$ vs. $v_w$ diagram can be used 
to determine if the interaction between the winds of SSCs are radiative, 
in which case cold filaments can be formed. 
With the $(\dot{M}_w/D)$ vs. $v_w$ diagram (Figure~1) one can
diagnose if a galaxy with a high star formation rate will show filaments 
in optical emission lines, or only a galactic wind with X-ray emission. 
At the same time, given an observation of a galactic wind with a rich 
filamentary structure, one could predict the number
of SSCs which are contained within the observed galaxy.

We computed three models of dense starbursts (M1-M3) based on the
observed parameters of M82, and a fourth model (M4), based on NGC253.
The models are based on cluster distributions with mean separations
between clusters similar to the observed values, but with spatial
extents which are considerably smaller than the total extent of
the starburst regions. The distributions used in the model
could correspond to individual ``starburst clumps'' such as the
ones observed in M82 by Westmoquette et al. (2007).

For the parameters of M82 we computed four models (M1-M3, and M1a)
with different metallicities (from solar to 10 times solar). For the
parameters of NGC253 we computed two models (M4 and M4a) with a 10 times
solar metallicity. Models M1-M4 asssumed a mass loss
rate consistent with the average mechanical luminosity of the
SSCs. Models M1a and M4a are obtained with rather extreme mass loss
rates, only achievable during a short-lived phase in which
the input from SN dominates over stellar winds.
 
For the models based on M82, we see the formation of cold,
H$\alpha$-emitting, filaments that extend out from the plane of the
the galactic disk in the models with metallicities $\gtrsim
5~Z_{\odot}$ (models M2 and M3). We also see formation of dense
filaments for the models with extreme mass loss rate at the supernova
phase of the clusters (models Ma1 and M4a), where a more
conservative mass loss rate did not yielded filaments (M1 and M4).

For the model based on NGC253 (M4), no filaments are produced,
even for a $10~Z_{\odot}$ metallicity.
This is consistent with the present observations of this object,
in which no H$\alpha$-emitting filaments have been detected.
The prediction obtained from our models would be that with the
distribution of SSCs of NGC253 (in pre or post SN phase), no H$\alpha$ 
filaments should exist. However, we have shown that during the SN phase
(model M4a and section~3.1)  the formation of cold filaments is 
possible in such high metallic  winds.  Hoopes et al. (2005) claim that 
the lack of evidence of filamentary structure in the galactic wind of
NGC253 could simply be the result of not having deep enough H$\alpha$
maps of this object. In the future we will see whether or not this
result is confirmed by deeper observations. 

We have also computed X-ray maps in the energy range of $0.3$
to $2~\rm{keV}$. For the highly radiative models (M2 and M3), we
find that the X-ray emission is concentrated in a region close to
the galactic plane. In order to reconcile this result with
the observation of extended X-ray emission in M82, we would have
to assume that the emission comes from another component in the flow,
which could be the shock of SGW against the intergalactic medium.

Finally, we have run two models (M2b and M3b) with the parameters
based on M82 (and abundances of $5~Z_{\odot}$ and $10~Z_{\odot}$,
respectively) with lower computational resolutions but with
larger spatial extents. From these runs, we find that the models
produce a filamentary H$\alpha$ morphology that extends out to
$\sim 0.5\to 1$~kpc on each side of the galactic plane.

We end by noting that in this paper we prove that the radiative
interaction between the winds from a disk-like distribution of
identical SSCs (super stellar clusters) does lead to the production of
dense filaments extending out to $100\to 1000$~pc away from
the galactic plane. These structures produce H$\alpha$ emission
that might correspond to the filaments observed in starburst
galaxies such as M82 and NGC1569. Of course, in our models
we do not consider many elements that are likely to be important
in starburst galaxies. Important among these might be:
\begin{itemize}
\item the presence of a dense, stratified
galactic disk interstellar medium,
\item clumpiness in this medium,
\item the existence of a distribution of sizes and ages for the SSCs,
\item the effect of galactic mass loading.

\end{itemize}
There is therefore a large field for future theoretical studies
on the formation of filamentary strutures in starburst
galaxies.

\begin{acknowledgements}
We acknowledge support from
the DGAPA (UNAM) grant IN108207, from the CONACyT grants 46828-F and
61547, and from the ``Macroproyecto de Tecnolog\'\i as para la Universidad
de la Informaci\'on y la Computaci\'on'' (Secretar\'\i a de Desarrollo
Institucional de la UNAM). We thank Enrique Palacios, Mart\'\i n Cruz and
Antonio Ram\'\i rez for their support of the servers in which the
simulations were carried out.
\end{acknowledgements}


%
\end{document}